\documentstyle[prl,aps,epsfig,multicol]{revtex}
\newcommand{\braced}[1]{\langle#1\rangle}
\begin {document} 

\title{Detecting flux creep in superconducting YBa$_2$Cu$_3$O$_{7-\delta}$ 
thin films via damping of the oscillations of a levitating permanent 
magnet.}

\author{R. Grosser, A. Martin, M. Niemetz, E.V. Pechen\cite{evp}, and W. 
Schoepe}
\address{Institut f\"ur Experimentelle und Angewandte Physik,
Universit\"at Regensburg,
D-93040 Regensburg,
Germany}

\date{\today}
\maketitle

\begin {abstract}
The damping of the oscillations of a small permanent magnet 
(spherical shape, radius 0.1 mm) levitating between two parallel epitaxial 
YBCO films is measured as a function of oscillation amplitude and
temperature. At small amplitudes the dissipation is found to be orders of 
magnitude lower than in bulk YBCO, Q-factors exceeding one million at low 
temperatures. With increasing amplitude the dissipation becomes 
exponentially large, exceeding the bulk values at large drives. We 
describe our results by calculating the ac shielding currents flowing 
through trapped flux whose motion gives rise to electric fields. 
We find dissipation to originate from different mechanisms of flux dynamics. 
\end{abstract}

\begin{multicols}{2}
A permanent magnet levitating above, or suspending below, a 
superconducting surface has an equilibrium position which is determined by 
magnetic forces due to screening currents and trapped flux 
\cite{rf1}. Except for a small logarithmic long-term relaxation of the 
magnetization of the superconductor this position is stable. The system is 
in a metastable state with stationary flux lines and consequently without 
any decay of the dc supercurrents, i.e., without any energy dissipation. 
However, oscillations of the levitating magnet about its equilibrium 
position give rise to ac magnetic fields at the surface of the 
superconductor which modulate the shielding currents. The initial state is 
now disturbed. The superconductor periodically tries to relax to more 
favorable metastable states by moving  flux lines  back and forth as a 
result of the oscillating component of the Lorentz forces. This leads to 
energy dissipation which we can determine from measurements of the 
oscillation amplitude of the magnet as a function of an 
external driving force.

In our present work we have investigated the energy 
dissipation when the magnet is levitating between two horizontal 
YBa$_2$Cu$_3$O$_{7-\delta}$ (YBCO) epitaxial thin films. We find that at 
small amplitudes and low temperatures the damping is several orders of 
magnitude smaller than with YBCO bulk samples studied in earlier work 
~\cite{rf3}, leading to Q-factors above 
$10^6$. This result may be of significance when application of 
superconducting levitation, e.g., for micromechanical bearings, is 
considered \cite{rf1}. 
At large driving forces and at temperatures above 77 K, however, the 
dissipation grows exponentially with the oscillation amplitude and 
ultimately even exceeds the bulk values. We discuss our results in terms 
of thermally activated flux flow phenomena in the superconducting films.

Our experimental method has been described in detail in earlier work on 
sintered YBCO \cite{rf3}, or niobium \cite{rf2}, and 
also when using the oscillating spherical magnet for hydrodynamic 
experiments in superfluid helium \cite{rf4}. It consists of placing a 
magnetic microsphere made of SmCo$_5$ (radius 0.1 mm) inside of a parallel 
plate capacitor (spacing $d = 1$ mm, diameter 4 mm) having electrodes which 
in our present work are made of YBCO epitaxial thin films (thickness 
$\delta = 450$ nm and in a second experiment 190 nm) 
laser deposited on insulating substrates \cite{rf5}. 
We prepared the YBCO films on SrTiO$_3$ (lower electrode) and 
Y$_2$O$_3$-stabilized ZrO$_2$ (upper electrode) substrates cut with an 
inclination of 2 degrees off the (001) orientation. The inclination 
improves both flux pinning and, due to initiating a terrace growth, 
crystallinity of the films. The lower electrode is protected by a 20 nm 
thick PrBa$_2$Cu$_3$O$_7$ epitaxial layer. The YBCO films have an 
extremely sharp superconducting transition at $T_c=89.5$~K measured by an 
ac susceptibility method, the complete transition width being 0.1 K.
Before the capacitor is cooled through $T_c$ we apply 
a dc voltage of about 800 V to the bottom electrode. Therefore, the magnet 
carries an electric charge $q$ of about 2~pC when levitating. The dc 
voltage is then switched off and oscillations of the magnet can be excited 
with an ac voltage $U_{ac}$ ranging typically from 0.1 mV to 20 V and 
having a frequency at the resonance of the oscillations ($\approx 300$ 
Hz). These vertical oscillations induce a current $qv/d$ in the 
electrodes, where $v$ is the velocity of the magnet, which is measured by 
an electrometer connected to a lock-in amplifier. For a given driving 
force $F = qU_{ac}/d$ we measure the 
maximum signal when slowly sweeping the frequency towards resonance. We then 
vary the driving force at constant temperature. In order to infer the 
velocity amplitude from the measured current amplitude and the driving force 
from the ac voltage we determine the charge $q$ by recording a resonance 
curve and analyzing its width (which is given by the damping coefficient) 
and its height which then determines the charge. The driving forces range 
from 10$^{-13}$ N to 10$^{-7}$ N, the velocity amplitudes from 0.1 mm/s 
to 50 mm/s, and the oscillation amplitudes $a = v/\omega$ from 50 nm to 30 
$\mu$m. The dissipated power varies from 10$^{-9}$ W down to below 
10$^{-17}$ W which is less than what is usually dissipated when 
investigating flux creep by measuring the ac susceptibility or the 
current-voltage characteristic of a type II superconductor.
\begin{figure}[t]
\centering\epsfig{file=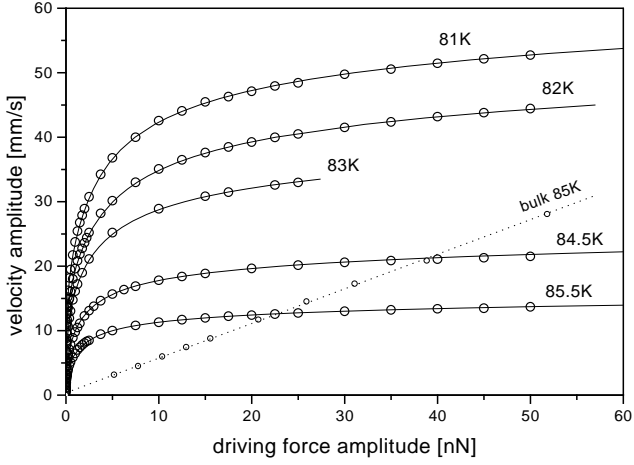,width=8.5cm, bb=80 280 735 765}
\begin{minipage}{8.6cm}\caption{Resonant velocity amplitude of the 
oscillating magnet as a 
function of the driving force at various temperatures. Note the steep 
increase at small driving forces and a logarithmic dependence at large 
drives. The lines are fits of our model to the data, see text. A 
typical linear dependence observed with melt-textured bulk samples is 
shown for comparison.}\end{minipage}
\label{fig1}
\end{figure}%
\begin{figure}[b]
\centering\epsfig{file=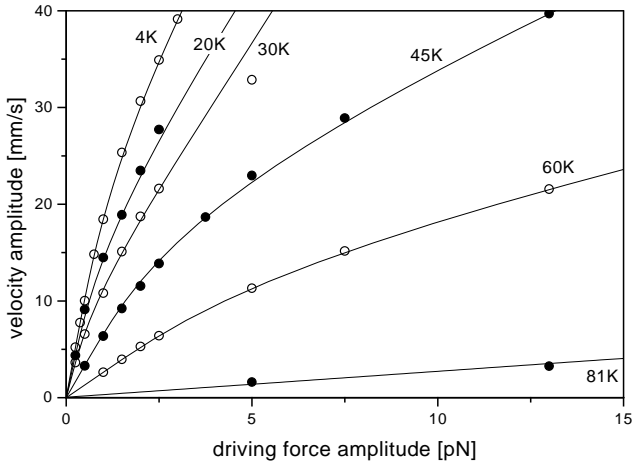,width=8.5cm, bb=80 280 735 765}
\begin{minipage}{8.6cm}\caption{Velocity amplitude at very small driving 
forces 
indicating the linear initial increase at various temperatures. 
The lines are fits of our model to the data, see text.}\end{minipage}
\label{fig2}
\end{figure}

A summary of our data obtained with the 450 nm films 
above 77~K is shown in Fig.~1. At small 
driving forces we find a steep initial increase of the velocity amplitude 
of the magnet, whereas at driving forces above ca.~10 nN there is only a 
logarithmic dependence. This implies an 
exponential increase of the dissipation. At lower 
temperatures this logarithmic regime could 
not be reached because the oscillations became unstable at large 
amplitudes. The small-signal behavior is shown in Fig.~2 with a
largely expanded abscissa. Only at very small amplitudes we do find a 
linear regime between the driving force amplitude $F$ and the 
velocity amplitude $v$, i.e., $F = \gamma v$ with a strongly temperature 
dependent coefficient $\gamma (T)$, see Fig.~3. While $\gamma$ 
decreases considerably below the critical temperature $T_c$, it 
obviously levels off at 4~K. Here the Q-factor $m \omega/\gamma$ ($m = 
5.2 \cdot 10^{-8}$ kg is the mass of the magnet) exceeds $10^6$~\cite{em}. 
The steep drop below $T_c$ follows approximately a $(T_c - T)^{-2}$ law as 
can be seen in the insert of Fig.~3. 
Comparing these results with our data on bulk YBCO \cite{rf3} we 
note that at 78 K $\gamma$ has decreased by three orders of magnitude.

For a more quantitative analysis we first calculate the sheet current on 
the surface of the superconductor by assuming that the magnetic field on 
either surface is given by the dipolar fields of the magnet and the first 
image dipole, 
neglecting images further away and also modifications of the field due to 
trapped flux lines. From the remanence of the magnet and its radius we 
calculate its magnetic moment to be $4 \cdot 10^{-6}$ Am$^2$. We further 
assume that the magnet levitates in the middle between the electrodes, i.e. 
at a height $h = h_0 = 0.5$ mm \cite{rf6}. It is simple to calculate the dc 
sheet current on the surface for a particular orientation of the dipole 
with respect to the surface \cite{rf7}. The orientation of the dipole 
parallel to the surface has the lowest energy but because of trapped flux 
we cannot rule out that some other orientation applies. Although the 
variation of the surface current $\vec{J}=(J_x(x,y,h_0),J_y(x,y,h_0))$ 
($x$ and $y$ are coordinates on
the surface) is quite complex and changes drastically with the orientation 
of the dipole, the maximum value $J_{max}$ is fairly insensitive to the 
orientation: $J_{max} = 4400$ A/m when the dipole is perpendicular and 
$J_{max} = 5100$ A/m for the parallel orientation. In either case the 
magnetic field is below $H_{c1}$, but some flux remains trapped because we 
have a field-cooled situation. 
We then calculate the ac sheet current 
$\Delta \vec{J}$  due to the oscillations of the magnet about its position 
at $h_0$:

\begin{equation}
\Delta \vec{J}(x,y,t) = \left(\left( \frac{\partial J_x}{\partial h}\right) 
_{h_0},\,\left( \frac{\partial J_y}{\partial h}\right) 
_{h_0}\right)
 a \cos(\omega t), \label{gleich1}
\end{equation}
where $a$ is the oscillation amplitude. Again the current distribution is 
complicated but the maximum amplitude of $\Delta J$ is insensitive to the 
orientation of the dipole. 
An oscillation amplitude $a = 0.5 ~\mu\mathrm{m}$ corresponding to a
velocity amplitude of $v \approx 1$ mm/s 
leads to a 
maximum current amplitude of $14$ A/m for the perpendicular orientation and 
of $15$ A/m for the parallel one. The current is localized in a small area 
of the superconducting surfaces below and above the sphere.
This ac sheet current is actually an 
oscillating current density $j$ in the film which exerts 
an oscillating Lorentz 
force on the trapped flux lines. Some flux lines will be set into motion 
thereby creating an 
electric field $E = v_{\rm f} n \Phi _0$, 
where $v_{\rm f}$ is the flux line velocity 
and $n$ is the number of flux lines per area. 
Because $j$ and $E$ are parallel the dissipation per unit volume is given by 
$j \cdot E$. If, for simplicity, we use an average current density 
$\braced{j}$ inside the film of thickness $\delta$, we can describe the dissipation 
per unit area by $\Delta J \cdot E$, where $E$ depends on 
$\braced{j} = \Delta J / \delta $ \cite{rf8}.
We can calculate the stationary oscillation amplitude at resonance by 
employing the energy balance between the gain from the driving force and the 
loss per cycle (period $\tau$):
\begin{equation}
\int_{0}^{\tau} F\cdot v \, dt = \int_{0}^{\tau}\!\!\!\int_{x,y} \Delta 
J\cdot 
E\, dx dy dt \label{gleich2}
\end{equation}
In the linear regime  $F=\gamma v$ the loss (r.h.s. of Eq. \ref{gleich2})
can be described by a surface resistance $R_s \propto \gamma$, namely 
\begin{equation}
\int_{0}^{\tau}\!\!\!\int_{x,y} 
R_s \cdot (\Delta J)^2 \, dx dy dt = R_s 
\int_{0}^{\tau}\!\!\!\int_{x,y} 
(\Delta J)^2 \, dx dy dt.
\label{gleich3}
\end{equation}
In Fig.~3 the 
$R_s$ values are given for the parallel orientation of the 
dipole \cite{error}.
From the surface resistance we can infer the imaginary part of the 
penetration depth of the ac field being given by 
$\lambda''=R_s/\mu_0\omega$.
In our case ($\omega/2\pi \approx 300 ~\mathrm{Hz}$) we find $\lambda''$ to 
drop from 84~nm at 85~K down to only 0.2~nm at 4~K. While at present we do 
not have a model for a quantitative analysis of these data, we note that a 
steep decrease of the surface resistance with temperature is also observed 
at both radio and microwave frequencies and is believed to be due to small 
oscillations of the trapped flux lines or, in zero field, due to normal 
fluid electrons.
\begin{figure}[t]
\centering\epsfig{file=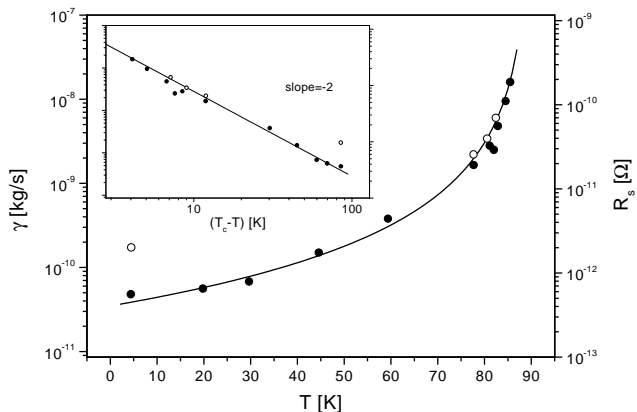,width=8.5cm, bb=50 280 810 780}
\begin{minipage}{8.6cm}\caption{Linear coefficient $\gamma = F / v$ 
from 
Fig.~2. Insert: same data (ordinate unchanged) plotted vs 
$T_c-T$. The surface resistance is calculated from Eq.~3.
Open symbols are data obtained with the thinner film 
$\delta=190\mathrm{nm}$}\end{minipage}
\label{fig3}
\end{figure}%
\begin{figure}[b]
\centering\epsfig{file=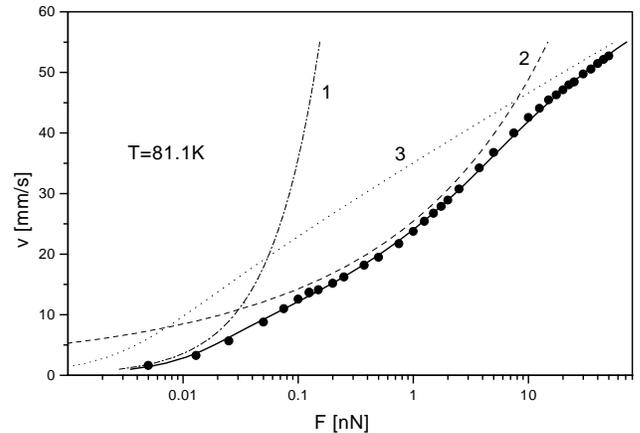,width=8.5cm, bb=60 280 765 765}
\begin{minipage}{8.6cm}\caption{Fit of Eq.~2 to a typical curve of Fig.~1. 
Indicated are 
the
separate contributions of three different mechanisms of dissipation: 1 is 
the linear one (which is an exponential curve because of the logarithmic 
abscissa), 2 is the contribution from Eq.~4, and  
3 is from Eq.~5. The solid line is the sum of these 
contributions.}
\label{fig4}\end{minipage}
\end{figure}

In the nonlinear regime of $v(F)$ the electric field in Eq.~\ref{gleich2} 
must have a nonlinear current dependence. This is usually described by 
thermally activated flux motion 
$v_{\rm f} \propto \exp(-U(j)/kT)$ with current 
dependent pinning energies $U(j)$, for a review see \cite{rf9}. 
Essentially  
two different dependences are being considered, firstly barriers diverging 
for $j \to 0$:
\begin{equation}
U(j) = U_0 \left( \frac{j_0}{j}\right)^\mu, \label{gleich4}
\end{equation}
where $j \le j_0$ and $0 < \mu < 1$. Vortex glass and collective creep 
models (for the latter ones $\mu > 1$ is also possible) yield this behavior 
which implies zero dissipation in the limit $j \to 0$. Secondly, for large 
currents, one considers barriers vanishing at a critical current density 
$j_c$ as 
\begin{equation}
U(j)=U_0 \left( 1- \frac{j}{j_c} \right),
\label{gleich5}
\end{equation}
which leads to an exponential dissipation for large currents (Kim-Anderson 
model), $E \propto \sinh(j/j_1)$ with $j_1= j_c kT / U_0$. We fit the 
complete $v(F)$ curves by including both nonlinear mechanisms (Eqs. 
\ref{gleich4}, \ref{gleich5}) and a linear term additively in the energy 
balance, see Eq.~\ref{gleich2}. An example is shown in Fig.~4, where the 
separate contributions and their sum are compared with an experimental 
curve.
The linear term contributes only at very low amplitudes. In the intermediate 
range the glass (or creep) term (Eq.~\ref{gleich4} with $\mu=0.2$) is 
dominant. Only at the largest amplitudes the Kim-Anderson term 
(Eq.~\ref{gleich5}) becomes relevant. It is this mechanism which leads to 
the observed slow logarithmic increase of the oscillation amplitudes at 
large drives in Fig.~1. Towards lower temperatures only the linear term and 
the glass term contribute as no data could be obtained at larger drives 
because the oscillations became unstable. From these fits we can determine 
the barriers $U(j)$ at various temperatures, see Fig.~5. It is a peculiarity 
of Eq.~\ref{gleich5} that only the constant slope $-dU/dj=U_0/j_c$ can be 
obtained reliably from a fit to the data because the quantity $U_0$ 
determines only the prefactor which we do not evaluate for it contains other 
unknown quantities, e.g. the density of the trapped flux $n$ or the 
prefactor of the vortex velocity $v_{\rm f}$. 
From Fig.~5 it is evident that at 
large currents and high temperatures the barriers of Eq.~\ref{gleich5} 
ultimately drop below those of Eq.~\ref{gleich4} \cite{rf10}. 
It is interesting to note 
that all three mechanisms of dissipation are found to apply simultaneously 
and additively. We do not find any indication of a sudden disappearance of 
one of the mechanisms which might be caused by a phase transition in the 
vortex system. In this context it may be important to remember that we are 
dealing with a low vortex density. We do find, however, that the critical 
current $j_c$ in Eq.~\ref{gleich5} extrapolates to zero at 86.5~K (well 
below $T_c=89.5$~K) which might indicate an onset of free flux flow 
where 
the magnet loses its lateral stability.
\begin{figure}
\centering\epsfig{file=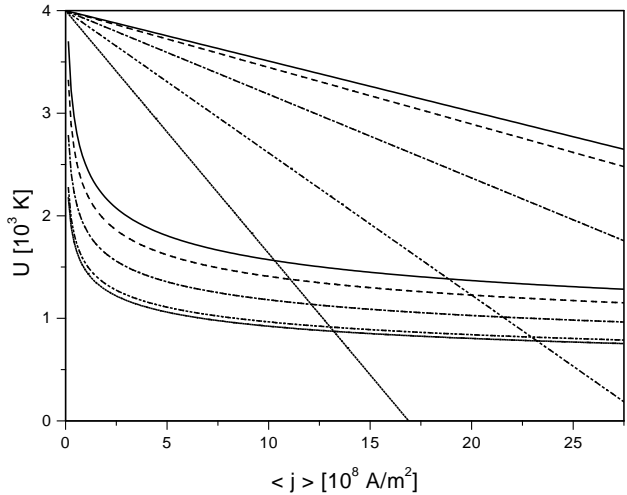,width=8.5cm, bb=90 270 725 775}
\begin{minipage}{8.6cm}\caption{Dependence of the pinning potential $U$ on 
the average current density in 
the film obtained from the fits of our model to the data: curved lines are 
Eq.~4, straight lines are Eq.~5 at the following 
temperatures (from top to bottom): 78~K, 81~K, 83~K, 84.5~K, 85.5~K.
Note that from Eq.~5 we can determine only the slopes $U_0/j_c$, 
in the figure we arbitrarily set $U_0=4 \cdot 10^3 \mbox{K}$
}\end{minipage}
\label{fig5}
\end{figure}

In summary, our method of studying the dynamic properties of a levitating 
magnet yields results on flux pinning which are supplementary to those of 
the more standard techniques like measurements of the ac susceptibility, 
magnetic relaxation, or current-voltage characteristic.  Because of 
the dipolar magnetic field the analysis of the results may be more 
complicated but our method has the advantage of being independent of 
edge effects and demagnetization factors. The 
enormous increase of the losses in the YBCO films from a very low level to 
an exponential dissipation is a surprise and is in sharp contrast to bulk 
samples. A more detailed account of our experiments including an analysis 
of the results on bulk samples (sintered and melt-textured) and of 
the elastic properties of the oscillating magnet will be presented 
elsewhere.

We are grateful to K.F. Renk for support of our work and to O. Kus for 
structuring the films. We had helpful discussions on vortex dynamics 
with N.B. Kopnin and E.B. Sonin. Financial support by 
the Deutsche Forschungsgemeinschaft is acknowledged by R.G. and M.N. 
(Graduiertenkolleg ``Komplexit\"at in Festk\"orpern'').

\end{multicols}
\end {document}